\documentclass[twocolumn,showpacs,amsmath,amstex,amssymb,mathfonts,aps,prl]{revtex4}
\usepackage{graphicx,epstopdf,verbatim}

\begin{document}
\title{Large Disorder Renormalization Group Study of the Anderson Model of Localization}
\author{Sonika Johri$^1$ and R. N. Bhatt$^{1,2}$}
\affiliation{$^1$ Department of Electrical Engineering, Princeton University, Princeton, NJ 08544}
\affiliation{$^2$ School of Natural Sciences, Institute for Advanced Study, Princeton, NJ 08540}

\begin{abstract}
We describe a large disorder renormalization group (LDRG) method for the Anderson model of localization in one dimension which decimates eigenstates based on the size of their wavefunctions rather than their energy. We show that our LDRG scheme flows to infinite disorder, and thus becomes asymptotically exact. We use it to obtain the disorder-averaged inverse participation ratio and density of states for the entire spectrum. A modified scheme is formulated for higher dimensions, which is found to be less efficient, but capable of improvement.
\end{abstract}
\date{\today}
\maketitle
Pioneering work on the application of renormalization group (RG) methods to highly disordered antiferromagnetic quantum spin systems in one dimension \cite{ma,bhatt_lee1} as well as higher dimensions \cite{bhatt_lee1, bhatt_lee2}  was initially greeted with skepticism. It appeared to be another real-space RG scheme, which was known to give poor results in uniform lattice models because of uncontrolled approximations, in contrast to Wilson's original k-space method for ($4-\epsilon$) dimensions \cite{wilson}. However, for disordered systems (unlike uniform systems), there exists a perfectly justified expansion parameter: the ratio of weak couplings to large couplings, or equivalently, the inverse width of the distribution of the {\it logarithm} of the couplings (see e.g. Fig. 1a of \cite{bhatt_lee2}); this provided a small parameter to allow a perturbative approach, and examine the RG flow.

While an analytic justification of the approach and the proof of its asymptotic exactness in one dimension took another decade and a half \cite{fisher}, numerically the result was already apparent in the early work (see e.g. Figure 7 of \cite{bhatt_lee1}). In the past two decades, the method of strong or large-disorder-renormalization group (LDRG) has been used to study several disorder models, especially in one dimension, including mixed antiferromagnetic-ferromagnetic couplings \cite{westerberg}, disordered binary chain \cite{hyman}, random transverse field Ising chain \cite{fisher2, refael1}, phase coupled oscillators \cite{kogan}, and bosons with strong disorder \cite{refael2}. A review summarizing some of these developments has appeared \cite{igloi}.

In this work, we apply the LDRG approach to the Anderson model of localization \cite{anderson}. Most aspects of the Anderson model, especially the localization transition, are well-known using techniques such as the non-linear sigma model as well as numerical approaches (see \cite{kramer_etc,evers,book} for a review of recent results). However, it was recently discovered numerically \cite{johriprl,johriprb} that the original model of Anderson with diagonal disorder described by a uniform bounded distribution, in the localized phase at moderate to high disorder, far from being featureless, exhibits a very abrupt, apparently singular, change in the nature of eigenstates as a function of energy. This feature arises due to a cross-over from typical Anderson localized states near the center of the band to a regime of resonant states near the edge of the band. It reveals itself in a sharp change in the first derivative of the density of states (DOS), and more prominently, the inverse participation ratio (IPR). Given the relatively few {\it quantitative} tests of LDRG, the Anderson model, by being a non-interacting model computable in polynomial time (i.e., essentially solvable numerically to very high precision), thus provides an ideal testing ground for a check on the accuracy of LDRG methods.

Here we formulate a LDRG scheme appropriate for studying the localized phase of the Anderson model. Our method is based on {\it eigenfunction} characteristics, in contrast to RG schemes for uniform systems based on {\it eigenvalues}. Because of the non-monotonic dependence of the spatial extent of wavefunctions with energy in the Anderson model, as well as the explicit use of the LDRG philosophy, our method differs from previous position-space RG studies of the Anderson model  ({\it e.g.} \cite{lambert, aoki, sarker, white, monthus}). For moderate to large disorder, our method accurately  captures the sharp change from typical Anderson localized states to resonant states (the latter eventually lead to the Lifshitz tail near the band edge). It provides accurate and quantitatively controllable results for the ensemble averaged density of states as well as the size of the wavefunctions {\it for the entire spectrum}. In one dimension, the scheme we propose flows to infinite disorder and thus errors remain controlled. Though the more approximate method we use in higher dimensions does not share such a simple flow, it still affords significant speed-up over exact diagonalization and sparse matrix methods, and further refinements are likely to yield greater accuracy.


The tight-binding Anderson model Hamiltonian on a $d$-dimensional hypercubic lattice \cite{anderson} is
\begin{equation}\label{eq:ham}
H_0 = \sum_{i}(\epsilon_i |i><i| + (V_{i,i+1}|i><i+1|	 +h.c..)	)				
\vspace{-10pt}
\end{equation}
where $|i>$ are (orthonormal) states localized on sites $i$ of a simple hypercubic lattice. The onsite energies $\epsilon_i$ are \textit{independent} random variables, with a distribution $P(\epsilon)$. As in the original Anderson work \cite{anderson}, we take the initial $P(\epsilon)$ to be a uniform distribution with width $w$, symmetric around $\epsilon = 0$ and set all nearest neighbour  hoppings $V_{i,i+1}=1$, which sets the overall energy scale. $w$ should be compared with the full bandwidth in the absence of disorder, which is $2Z$, where $Z$ is the coordination number of the lattice.  We define $x = w/(2Z)$ which is thus a measure of the disorder strength in the system. We use periodic boundary conditions. However, the RG method outlined below can be applied for all boundary conditions and initial probability distributions. During the course of the RG, the distributions of both $\epsilon$ and $V$ will be modified.

The IPR for a wavefunction $\Psi=\sum_{i}a_i |i>$ is defined as 
\vspace{-5pt}
\begin{equation}\label{eq:defipr}
I_{\psi}=\frac{\sum_{i} {|a_i|^4}}{(\sum_{i}{|a_i|^2})^2}
\vspace{-5pt}
\end{equation}
$I_{\psi}$ is thus inversely proportional to the number of sites where the wavefunction has significant amplitude. 

The basic idea of our LDRG scheme is to extract eigenstates from the system starting with the {\it most localized} ones, irrespective of the energy of the eigenstate. 
For zero hopping (or equivalently infinite disorder $w$), all eigenstates are restricted to one site only. For large $w$, we calculate the size of the eigenstate wavefunctions to first order using perturbation theory. We define the effective ``bond'' between the states at $i_0$ and $i_0+1$ by:
\begin{equation}
m_{i_0,i_0+1}=\frac{V_{i_0,i_0+1}}{E_{i_0}-E_{i_0+1}}.
\end{equation}
For small bonds $m$, the perturbed state at site $i_0$ has wavefunction and energy given by:
\begin{eqnarray}\label{eq:one_diag}
\nonumber\Psi_{i_0}'&=&|i_0>+m_{i_0,i_0+1}|i_0+1>-m_{i_0-1,i_0}|i_0-1>\\
d_{i_0}'&=&E_{i_0}+m_{i_0,i_0+1}V_{i_0,i_0+1}-m_{i_0-1,i_0}V_{i_0-1,i_0}
\end{eqnarray}
giving rise to an IPR 
\begin{equation}\label{eq:one_ipr}
I_{i_0}=\frac{1+m_{i_0-1,i_0}^4++m_{i_0,i_0+1}^4}{(1+m_{i_0-1,i_0}^2++m_{i_0,i_0+1}^2)^2}
\end{equation}
Since IPR is inversely proportional to the size of the wavefunction, and we want to flow in the direction of increasing wavefunction size, we start the RG with the site with the highest $I_i$. Since $I_{i}=1-2(m^2_{i,i+1}+m^2_{i-1,i})+\mathcal{O}(m^4)$ for small $m$, we use as our  RG flow parameter $m=\min(m^2_{i,i+1}+m^2_{i-1,i})$.

We remove and store the wavefunction and energy obtained in Eq. \ref{eq:one_diag}. The lattice now has one less site. We also renormalize the energies and wavefunctions of the erstwhile neighbours of $|i_0>$. The left neighbour of $|i_0>$ is perturbed as:
\begin{eqnarray}
\nonumber\Psi_L&=&|i_0-1>+m_{i_0-1,i_0}|i_0>\\
E_L&=&E_{i_0-1}+m_{i_0-1,i_0}V_{i_0-1,i_0},
\end{eqnarray}
and similarly for the right neighbour $\Psi_R$. The perturbation theory generates a hopping between these renormalized states, $V_{LR}=<\psi_L|H|\psi_R>$, where $H$ is the modified Hamiltonian at this stage of the LDRG. With these new values, we also recalculate the bonds for the nearest and next-nearest neighbors of the removed site. We repeat this procedure for all sites with both bonds less than $m_0$, flowing in the direction of increasing $m$. $m_0$ is a cut-off which should be smaller than $1$. The smaller the value of $m_0$, the more accurate the energies and wavefunctions obtained by this method.

Once there are no more sites with both bonds less than $m_0$ left in the lattice, we start removing ``2-site'' clusters in a similar fashion. Once all the 2-site clusters are finished, we remove the 3-site ones and so on. The procedure to remove an $n$-site cluster is similar in spirit to the one for a single site. Consider the cluster which consists of sites from $i_0$ to $i_0+n-1$ where $m_{i_0-1,i_0}$ and $m_{i_0+n-1,i_0+n}$ are smaller than $m_0$, and all other bonds in between are greater than $m_0$. We diagonalize the ``cluster-Hamiltonian'': $H^{(i_0)}= \sum_{i=i_0}^{i_0+n-2} (\epsilon_i |i><i|+(V_{i,i+1} |i><i+1|+h.c.)) +\epsilon_{i_0+n-1} |i_0+n-1><i_0+n-1|$ to give eigenstates $\Psi^{(i_0)}_j=\sum_{k=i_0}^{i_0+n-1} c^{(i_0,j)}_k|k>$ with corresponding eigenvalues $d^{(i_0)}_j$, where $j$ goes from $1$ to $n$. Perturbation of these wavefunctions with the rest of the lattice gives:
\begin{eqnarray}\label{eq:cldiag}
\nonumber \Psi'^{(i_0)}_j&=&\Psi^{(i_0)}_j-m_{i_0-1,i_0}c^{(i_0,j)}_{i_0}|i_0-1>\\
\nonumber&+&m_{i_0+n-1,i_0+n}c^{(i_0,j)}_{i_0+n-1}|i_0+n>\\
\nonumber d'^{(i_0)}_j&=&d^{(i_0)}_j-m_{i_0-1,i_0}V_{i_0-1,i_0}c^{(i_0,j)}_{i_0}\\
&+& m_{i_0+n-1,i_0+n}V_{i_0+n-1,i_0+n}c^{(i_0,j)}_{i_0+n-1}
\end{eqnarray}
We remove and store these $n$ energies and wavefunctions. The number of sites decreases by $n$ after this step. The site that was to the immediate left of the cluster is now perturbed as:
\begin{eqnarray}
\nonumber\Psi_L&=&|i_0-1>+m_{i_0-1,i_0}\bigg(\sum_{j=1}^{n} c^{(i_0,j)}_{i_0}\bigg)|i_0>\\
E_L&=&E_{i_0-1}+m_{i_0-1,i_0}V_{i_0-1,i_0}\bigg(\sum_{j=1}^{n} c^{(i_0,j)}_{i_0} \bigg)
\end{eqnarray}
and similarly for the site to the immediate right. A hopping is generated between $\Psi_L$ and $\Psi_R$ like in the one-site case. In order to select which cluster to remove first, we calculate a quantity similar to that in Eq. \ref{eq:one_ipr}, with $m_{i_0,i_0+1}$ replaced by $m_{i_0+n-1,i_0+n}$.

In one dimension, the LDRG does not destroy the lattice structure, i.e. each site always has two nearest neighbours. However, the states at each site may become combinations of several of the original tight-binding states, and the basis is no longer orthonormal. Thus when each cluster is diagonalized as in Eq. \ref{eq:cldiag}, a generalized eigenvalue equation has to be solved.

\begin{figure}
\includegraphics[width=\columnwidth]{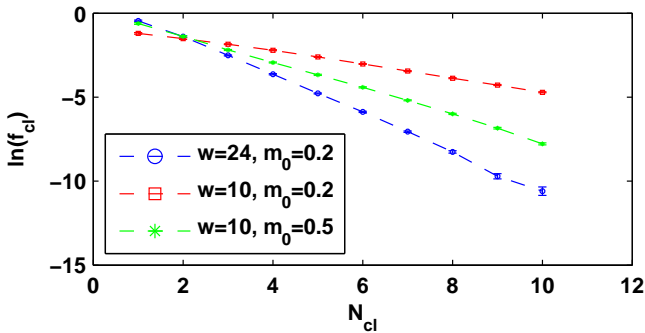}
\vspace{-20pt}
\caption{The fraction of the total system size, $f_{cl}$ that is decimated as a function of cluster size $N_{cl}$ during the LDRG flow for the one dimensional Anderson model with initial system size $L=10^5$. LDRG data is averaged over $100$ runs.}
\vspace{-15pt}
\label{fig:cl_size_length}
\end{figure}
Fig. \ref{fig:cl_size_length} plots the number of sites removed during the N-site cluster decimation process, $f_{cl}(N)=(L_{N-1}-L_{N})/L_0$, where $L_N$ is the number of sites remaining in the system after all clusters of size $N$ have been decimated. Since the definition of a cluster depends on the bond cut-off $m_0$, the number of large clusters is smaller for larger $m_0$. Large disorder also results in smaller clusters. An exponential decay of $f_{cl}$ with $N_{cl}$ is a consequence of independent random on-site energies (i.e. a Poisson distribution); this basic dependence appears to be retained within our LDRG scheme.

\begin{figure}
\includegraphics[width=\columnwidth]{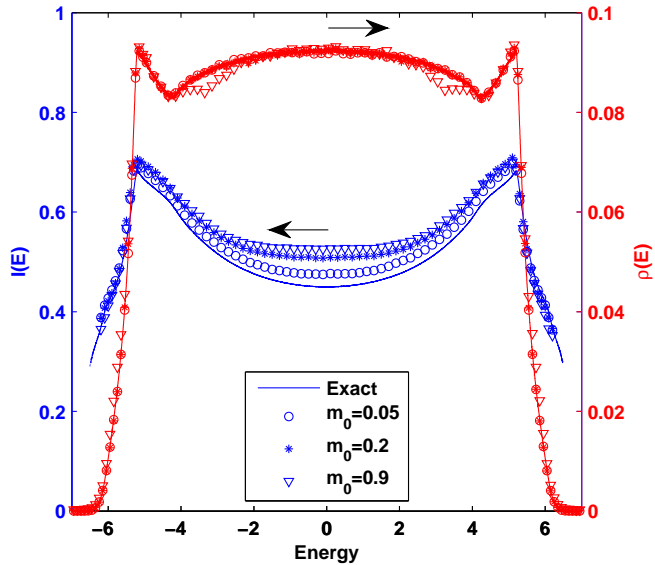}
\vspace{-20pt}
\caption{IPR (in blue, left y-axis) and DOS (in red, right y-axis) for the Anderson model in one dimension with $x=2.5$ $(w=10)$ from exact diagonalization (solid lines) and from LDRG with initial system size $L=10^5$, using different values of the cut-off $m_0$. LDRG data is averaged over $100$ runs.}
\vspace{-15pt}
\label{fig:ipr}
\end{figure}
Fig. \ref{fig:ipr} compares the disorder-averaged IPR and DOS for $w=10$ in one dimension from exact diagonalization (ED) and LDRG. Results from different values of the cutoff $m_0$ are shown. The accuracy decreases as $m_0$ is increased, because higher-order perturbations become more significant when bonds are stronger. 
It may be seen that the resonant states leading to the Lifshitz tail are well captured by the LDRG. This is because the resonant states are in fact a set of strongly coupled sites, which are loosely coupled to the rest of the system, i.e. the clusters in our LDRG. Resonant states composed of sites with energies close to the disorder edge ($w/2$) give rise to states with energies $|E| > w/2$, i.e towards the band edge. We remark, however, that most clusters are not resonant states, and have energies all across the spectrum.

To determine the accuracy of the LDRG, we evaluate the difference between its results and those obtained by exact diagonalization. For the DOS, we find that that the average error across the entire band at $w= 12$ is $ 0.1 \%$ for $m_0 = 0.2$, and significantly lower for $m_0 = 0.05$. Both errors decrease monotonically as $w$ increases. For the average IPR,
we define the following measure of the error within a given interval $(E_1, E_2)$
\begin{equation}
\delta I_{E_1,E_2}=\frac{1}{E_2 - E_1}\int_{E_1}^{E_2} |I_{RG}(E)-I_{ED}(E)| dE
\end{equation}
where $I_{RG}$ and $I_{ED}$ are the IPR obtained using the RG scheme and exact diagonalization respectively.

\begin{figure}
\includegraphics[width=\columnwidth]{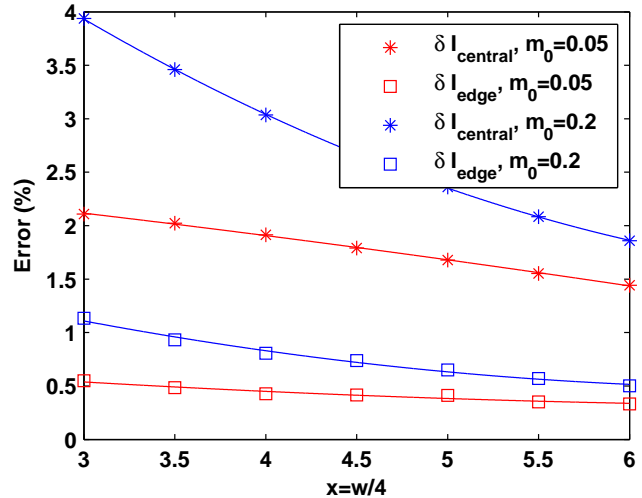}
\vspace{-20pt}
\caption{The error in the IPR (as defined in the text) obtained using LDRG with initial system size $L=10^5$ as a function of disorder, $x=w/4$ for the one dimensional Anderson model. Line is a guide to the eye.  LDRG data is averaged over $100$ runs. A similar measure can be defined for the DOS and lies below $0.1 \%$ for the values of $w$ shown here.}
\vspace{-15pt}
\label{fig:avg_ipr_vs_w}
\end{figure}

We divide the band into two parts - the main central portion, corresponding to $E_1 = - w/2$ and $E_2 = + w/2$, and the edge of the band, defined by $E_1 = w/2$ and $E_2 = w/2 + 2d$, plus the corresponding particle-hole conjugate $E_1 = -w/2 - 2d$ and $E_2 = - w/2$. We denote the average errors for the two regions by $\delta I_{\text{central}}$ and $\delta I_{\text{edge}}$.
Fig. \ref{fig:avg_ipr_vs_w} plots $\delta I_{\text{central}}$ and $\delta I_{\text{edge}}$ as a function of disorder $w$ for different values of $m_0$. (In practice $\delta I_{\text{edge}}$ is cut off a little before $w/2+2d$ when there is insufficient data due to the low density of states near the band-edge.) As can be clearly seen, $\delta I_{\text{edge}}$ is smaller; however both errors decrease as $w$ increases, and as $m_0$ decreases, clearly delineating the path for increased accuracy. 

\begin{figure}
\includegraphics[width=\columnwidth]{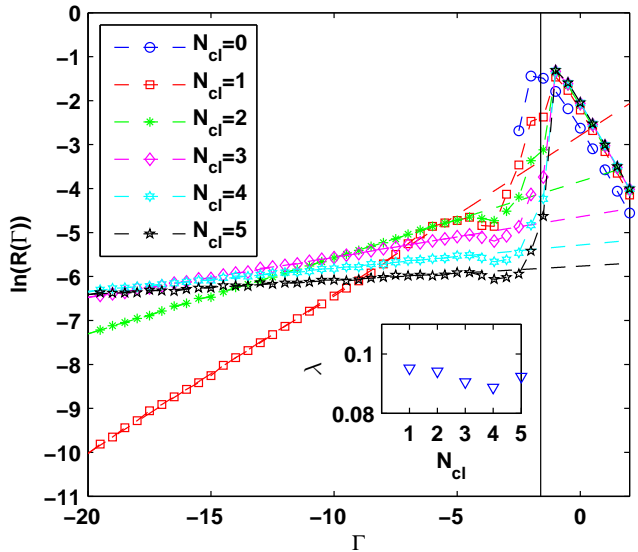}
\vspace{-20pt}
\caption{Evolution of the distribution of the logarithms of bond values $m_{ij}$ as the LDRG progresses for the one dimensional Anderson model with $x=2.5$ $(w=10)$ and $m_0=0.2$. Initial system size is $L=10^5$. $N_{cl}$ is the size of the clusters decimated just before the distribution was measured. $N_{cl}=0$ is thus the initial distribution.  LDRG data is averaged over $100$ runs. The vertical line marks $\Gamma_0$.  The straight lines are fits to $\ln(R(\Gamma))=b\Gamma+\ln(a)$. The inset shows that $\lambda=\frac{a}{b}e^{\Gamma_0 b}$ is approximately constant.  }
\vspace{-15pt}
\label{fig:mixwfn}
\end{figure}
\begin{figure}
\includegraphics[width=\columnwidth]{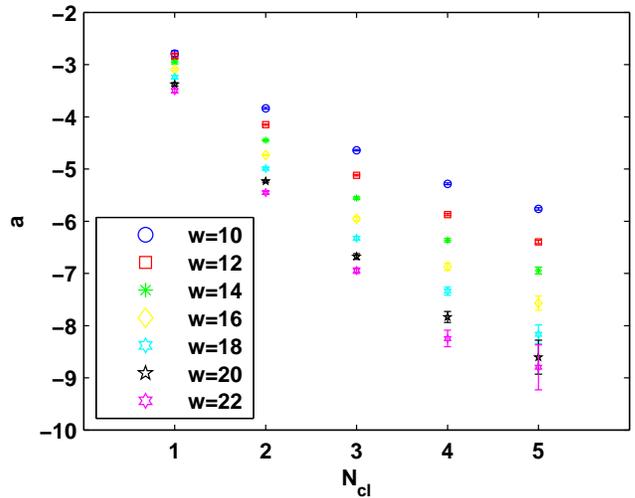}
\vspace{-20pt}
\caption{The evolution of the fitting parameter $a$ as a function of cluster size $N_{cl}$ for different values of the disorder $w$ and cut-off $m_0=0.2$. The error bars are $95\%$ confidence bounds of the fit.}
\vspace{-15pt}
\label{fig:mixwfn_fit_params}
\end{figure}

Since new bonds are generated during the RG flow by multiplying decimated bonds, it is convenient to consider the logarithmic variable $\Gamma_{ij}=\ln(m_{ij})$. Fig. \ref{fig:mixwfn} shows the evolution of of the distribution of this variable, $R(\Gamma)$, as the RG progresses. The initial distribution is the blue curve. The distributions can clearly be divided into two parts on either side of $\Gamma=\Gamma_0=\ln(m_0)$. $R(\Gamma>\Gamma_0)$ does not change as the RG progresses, because the probability of any of the strong bonds being removed inside a cluster is equal. $R(\Gamma<\Gamma_0)$ can be fitted by an exponential of the form $a\exp(b\Gamma)$, i.e. straight lines in Fig. \ref{fig:mixwfn}. This implies that the $m_{ij}$ have a power-law probability distribution given by $F(m)=am^{b-1}$ \cite{hyman}. $a$ and $b$ are related since the integral over the probability distribution is equal to 1. Approximately, we can see that $\frac{a}{b}e^{m_0 b}=\lambda$, where $\lambda$ is some constant which does not change during the RG flow. The inset in Fig. \ref{fig:mixwfn} shows that this is true within error bounds of the fitting estimate for $a$ and $b$. Thus, the width of the distribution is given by $1/a$. Fig. \ref{fig:mixwfn_fit_params} shows that the parameter $a$ decreases as the LDRG evolves, showing that the RG flows to infinite disorder. This again strongly suggests that within this scheme, which is much faster than standard diagonalization, and applicable to much larger system sizes, there is a systematic method for decreasing errors in (at least) disorder-averaged quantities related to eigenvalues and eigenfunctions. For two such quantities, the density of states and the inverse participation ratio, the scheme can be implemented practically down to sufficient accuracy as to capture their salient features, for reasonably high disorder.


We now turn to higher dimensions. The RG scheme implemented in one dimension does not generate additional couplings, and leads to a convergent result as a consequence of flow to increasing disorder. In higher dimensions, the topology of the lattice changes under the RG, and leads to large connectivity with increasing complexity. The number of nearest neighbors is not fixed and a systematic delineation into clusters is unclear. The perturbative approach also breaks down because of constructively interfering paths. However, for large disorder values, we are still successful in obtaining IPR and DOS using a modified approach described below, which should in principle be applicable in any dimension. We perform two kinds of decimation:

1. Site Decimation: We remove single sites by the same method as in 1D, with the modification that each site can now have any number of bonds.

2. Bond Decimation: We eliminate bonds $m_{ij}$ larger than $m_0$ by diagonalizing the $2\times 2$ matrix $H_{ij}=\epsilon_i |i><i| +\epsilon_j |j><j| + V_{ij}|i><j|+h.c.$. This will change the basis and generate extra bonds, which may be weaker. We set a floor, $m_{\min}$ on the minimum value of stored bonds, and do not decimate bonds that have already been affected by a rotation. This ensures that rotations are independent from each other and effective at removing only the largest bonds. The procedure is continued till the total number of bonds stored is equal to a maximum $N_m$.

Rotations increase the number of bonds stored, and therefore increase both the memory consumption and the time required during each decimation step. Therefore, we exit step 2 and restart step 1 when the total number of bonds stored becomes greater than $N_m$, and place a cut-off on the minimum value of a bond $m_{min}$. 

\begin{figure}
\includegraphics[width=\columnwidth]{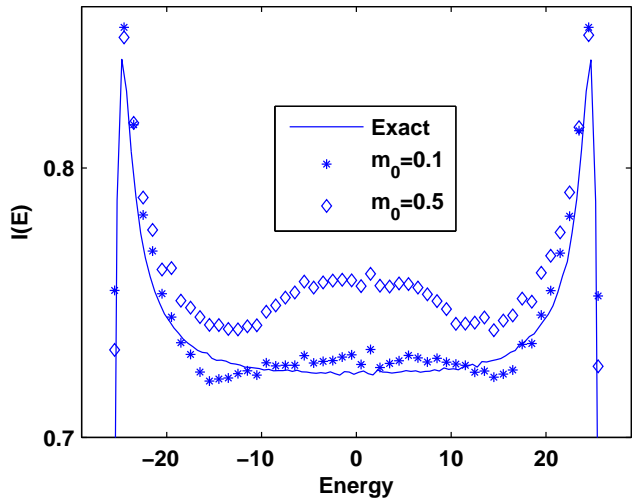}
\vspace{-20pt}
\caption{IPR for Anderson model in two dimensions with $x=6.25$ $(w=50)$ from exact diagonalization (solid line) and from LDRG with different values of the cutoff $m_0$. LDRG data is averaged over $100$ runs of systems with $100\times 100$ sites. }
\vspace{-15pt}
\label{fig:2d_ipr_dos}
\end{figure}
Fig. \ref{fig:2d_ipr_dos} compares IPR and DOS from exact diagonalization (ED) and this scheme for the square lattice in two dimensions for $x=6.25$ $(w=50)$ with $m_{\min}=10^{-4}$. As in the one-dimensional case, smaller values of $m_0$ produce more accurate results. In our data, we observed that for this value of disorder, the size of the lattice could be reduced by 90\%  before there was a significant increase in the number of bonds. A practical method to obtain information about wavefunctions of such a system would then be to utilize the RG to reduce the system to sizes where ED could work. 


In conclusion, we have proposed and implemented a LDRG scheme for the Anderson model of localization based on wavefunction size rather than any energy scale. This LDRG is especially useful when length and energy scales are not monotonically related, as is often true in disordered localized systems. The method provides access to essentially all eigenstates and eigenvalues of the system computed approximately using the perturbative RG approach. While more approximate, this method is significantly faster than either exact numerical diagonalization, or sparse matrix methods of diagonalization (which has to be performed repeatedly for computing quantities across the spectrum). Further, by using the perturbative parameter as a control, we are able to reduce errors and provide quite accurate results for ensemble averaged quantities such as the density of states and the inverse participation ratio at moderately high disorder in the localized phase. In one dimension, the RG is controlled as it flows to large disorder, and the scheme becomes more accurate as the RG proceeds. In higher dimensions, we use a modified approach which reduces the size of the system to a small fraction, at which point exact diagonalization may become feasible. Our method may also be useful for other problems such as many-body localization where the ``size'' of the wavefunction is measured in Fock space.

This work was supported by DOE grant DE-SC0002140. R. N. B. acknowledges the hospitality of the Institute for Advanced Study during the period when this work was completed and paper written. S. J. was supported by the Porter Ogden Jacobus Fellowship of Princeton University.

\end{document}